\documentclass[a4paper, 12pt, oneside]{article}
\usepackage{jheppub}
\makeatletter
\def\@fpheader{\hfill\today}
\makeatother
\usepackage{amsmath,graphicx,color,epsfig,amssymb,float}
\usepackage[normalem]{ulem}
\usepackage{epstopdf}
\newcommand{\nc}{\newcommand}
\nc{\non}{\nonumber}
\nc{\hc}{\hbox {H.c.}}
\nc{\noi}{\noindent}
\nc{\barx}{\bar{x}}
\nc{\pbarn}{\;\hbox {pb}}
\nc{\fbarn}{\;\hbox {fb}}
\nc{\hsp}{\hspace{0.5cm}}
\nc{\lsp}{\hspace{1cm}}
\nc{\Lsp}{\hspace{2cm}}
\nc{\LLsp}{\hspace{3cm}}
\nc{\lra}{\longrightarrow}
\nc{\p}{\prime}
\nc{\sgn}{\text{sgn}}

\nc{\beq}{\begin{equation}}  \nc{\eeq}{\end{equation}}
\nc{\bea}{\begin{eqnarray}}  \nc{\eea}{\end{eqnarray}}
\nc{\baa}{\begin{array}}     \nc{\eaa}{\end{array}}
\nc{\bit}{\begin{itemize}}   \nc{\eit}{\end{itemize}}
\nc{\ben}{\begin{enumerate}} \nc{\een}{\end{enumerate}}
\nc{\bce}{\begin{center}}    \nc{\ece}{\end{center}}
\nc{\bpm}{\begin{pmatrix}}   \nc{\epm}{\end{pmatrix}}
\nc{\bvt}{\begin{verbatim}}  \nc{\evt}{\end{verbatim}}

\def\lsim{\mathrel{\raise.3ex\hbox{$<$\kern-.75em\lower1ex\hbox{$\sim$}}}}
\def\gsim{\mathrel{\raise.3ex\hbox{$>$\kern-.75em\lower1ex\hbox{$\sim$}}}}

\def\udots{\mathinner{\mkern1mu\raise1pt\vbox{\kern7pt\hbox{.}}\mkern2mu\raise4pt\hbox{.}\mkern2mu\raise7pt\hbox{.}\mkern1mu}}

\def\gev{\;\hbox{GeV}}

\title{Generalized Randall-Sundrum model with a single thick brane}
\author[\ast]{Aqeel Ahmed,\note[$\ast$]{On leave of absence from National Centre for Physics,
Quaid-i-Azam University Campus, 45320 Islamabad, Pakistan}}
\author[]{Lukasz Dulny}
\author[]{and Bohdan Grzadkowski}
\affiliation[]{Faculty of Physics,
University of Warsaw,\\
Ho\.za 69, 00-681 Warsaw, Poland}
\emailAdd{aqeel.ahmed@fuw.edu.pl}
\emailAdd{lukasz.dulny@fuw.edu.pl}
\emailAdd{bohdan.grzadkowski@fuw.edu.pl}

\abstract{
A generalized version of the Randall-Sundrum model-2 with different
cosmological constants on each side of a brane has been discussed. A possibility
of replacing the singular brane by a configuration of a scalar field has been also
considered, the Einstein equations for this setup were solved and stability of the
solution discussed. It has been shown that under mild assumptions the relation
between cosmological constants and the brane tension obtained in the brane limit
does not depend on the particular choice of the regularizing profile of the scalar field.
}

\arxivnumber{1312.3577}

\begin{document}
\maketitle
\flushbottom


\section{Introduction}
\label{introduction}

The idea of extra dimensions offers a possibility of explaining
the hierarchy between the Planck scale $M_{Pl}\simeq 10^{18}\gev$ and
the electroweak scale $m_W\simeq 10^2\gev$, therefore it has received a lot of
attention during last decade. Randall and Sundrum proposed a very
elegant model (RS1) to solve the hierarchy
problem \cite{Randall:1999ee} and also an attractive alternative (RS2) for a
compactification of the extra dimension \cite{Randall:1999vf}. Both models
suffer from the presence of infinitesimally thin structures, so-called D3 branes.
In addition the RS1 requires the presence of a brane with negative tension.
There were many attempts to regularize thin branes of RS1 by certain configurations of a
scalar field with localized energy density. Unfortunately, it turns out
that periodicity constraints the dynamics of those models so strongly that
only trivial (constant) configurations of the scalar field are allowed, see
\cite{Gibbons:2000tf} and \cite{Ahmed:2012nh}. Therefore, here we are going to
limit ourself to the case of uncompactified extra dimension, \`a la RS2.
We will consider a generalized version of the RS2 allowing for different cosmological
constants on both sides of the brane. In this case a nontrivial profile
of the scalar field is allowed and a thick (smooth) brane could be adopted
to regularize the singular thin brane.
There have been many studies devoted to thick branes with different motivations
and setups~\cite{Ahmed:2012nh,DeWolfe:1999cp,Kehagias:2000au,Rubakov:1983bb,Gremm:1999pj,
Csaki:2000fc,Kobayashi:2001jd,Melfo:2002wd,Bronnikov:2003gg,
Bazeia:2003aw,CastilloFelisola:2004eg,Guerrero:2005aw,Farakos:2007ua,Bazeia:2008zx,Guo:2010az}, for review see for example \cite{Dzhunushaliev:2009va} and references therein.
In order to obtain a desired (warped) form
of solutions for the Einstein equations, both in the RS1 and the RS2 one has to impose
certain relations between the brane tension and cosmological constants.
Here we are going to prove that under certain mild assumptions,
the relation  between the brane tension and the cosmological constants obtained
in the brane limit of the thick brane scenario
does not depend on detailed shape of the scalar field profile.

The paper is organized as follows. The generalized RS2 model is defined in
Sec.~\ref{RS2 generalization}. Section~\ref{Thick brane version of RS2 generalization} contains
discussion of the thick brane version of the generalized RS2.
In Sec.~\ref{Generalized thick branes} we show that the RS2 relation between
the brane tension $\lambda$ and cosmological constants $\Lambda_\pm$ does not depend on
details of the thick brane profile. Section~\ref{sum} summarizes our findings.


\section{RS2 generalization}
\label{RS2 generalization}

We will consider the following action which is an extension of the Randall-Sundrum model with a
single brane (RS2) \cite{Randall:1999vf}~\footnote{When our work was
completed we came across the paper by Gabadadze et al. \cite{Gabadadze:2006jm},
where the authors also discussed the generalised RS2 (thin brane) model in detail.
Therefore here we summarize only those important aspects of the asymmetric model
that are necessary for the remaining parts of this paper.},
\beq
S=\int d^5x\sqrt{-g}\left(2M_\ast^{3}R-\Lambda_+\Theta(y-y_0)-\Lambda_-\Theta(-y+y_0)-\lambda\delta(y-y_0)\right),  \label{RS_acion}
\eeq
where $\Lambda_+$ and $\Lambda_-$ are 5D cosmological constants for $y>y_0$ and $y<y_0$, respectively,
whereas, $y_0$ is the brane location and $\lambda$ represents the brane tension. In the above action
$M_\ast$ is the 5D Planck mass. In our convention capital Roman indices will refer to 5D objects,
i.e., $M,N,\cdots=0,1,2,3,5$ while the Greek indices will label four-dimensional (4D) objects, i.e.,
$\mu,\nu,\cdots=0,1,2,3$. In Eq. \eqref{RS_acion} $\Theta$ is the Heaviside theta function and
$\delta$ is the Dirac delta function. For simplicity we will choose $y_0=0$.

We are going to look for solutions of the Einstein equations assuming the following form of the 5D metric
\beq
ds^2=e^{2A(y)}\eta_{\mu\nu}dx^\mu dx^\nu+dy^2.  \label{metric}
\eeq
Then the Einstein equations following from the action \eqref{RS_acion} reduce to,
\bea
6A^{\p2}&=&-\frac{1}{4M_\ast^{3}}\left(\Lambda_+\Theta(y)+\Lambda_-\Theta(-y)\right),  \label{eeq_55}\\
3A^{\p\p}+6A^{\p2}&=&-\frac{1}{4M_\ast^{3}}\left(\Lambda_+\Theta(y)+\Lambda_-\Theta(-y)+\lambda\delta(y)\right),    \label{eeq_mn}
\eea
The solution of Eq. \eqref{eeq_55} is given by,
\beq
A(y)= - |y| k_\pm \lsp {\rm for} \lsp y\gtrless 0,    \label{A_RS2}
\eeq
where $ k_\pm\equiv\sqrt{-\frac{1}{24M_\ast^{3}}\Lambda_\pm}$ can be related to the AdS curvatures
$R_\pm$ for $y\gtrless 0$ as $ k_\pm\propto 1/R_\pm$. Now one can calculate the $A^\p$ and $A^{\p\p}$
from the above expression as,
\beq
A^\p(y)= \mp k_\pm \;\; {\rm for} \;\; y\gtrless 0
\lsp {\rm and} \lsp  A^{\p\p}(y)=-( k_++ k_-)\delta(y).
\label{Ap_RS2}
\eeq
Discontinuity of $A^\p(y)$ at $y=0$ results in the following jump
\beq
\left[A^\p\right]_0=-\frac{\lambda}{12M_\ast^{3}}, \label{Ajump}
\eeq
where $[A^\p]_0\equiv A^\p(0+\epsilon)-A^\p(0-\epsilon)$, where $\epsilon\to 0$. From Einstein
equations \eqref{eeq_55} and \eqref{eeq_mn}, we have,
\beq
A^{\p\p}(y)=-\frac{\lambda}{12M_\ast^{3}}\delta(y).    \label{App_RS2_1}
\eeq
Comparing \eqref{App_RS2_1} and the second equation of \eqref{Ap_RS2} yields,
\beq
\lambda=\sqrt{6 M_\ast^{3}}\left(\sqrt{-\Lambda_+}+\sqrt{-\Lambda_-}\right),
\label{App_RS}
\eeq
which is an analogue of the Randall-Sundrum relation between the bulk cosmological constant and
the brane tension \cite{Randall:1999ee,Randall:1999vf}. It is important to note that the relation
\eqref{App_RS} is necessary in order to recover the 4D Poincar\'e invariance on the brane.

As we have checked by explicit calculation the 4D effective gravity on the brane could be recovered with the Planck mass given by:
\beq
M^{2}_{Pl}=\frac{M_\ast^3}{2k_+}+\frac{M_\ast^3}{2k_-},
\label{4D_Planck mass}
\eeq
We have also verified that the above solutions of the Einstein equations are stable against small perturbations of the metric. Our findings concerning the asymmetric version of the RS2 with singular brane confirm results obtained in \cite{Gabadadze:2006jm}. Some of the aspects of asymmetric singular brane worlds are discussed in \cite{Gregory:2000jc,Dvali:2000hr,Dvali:2000rv,Csaki:2000pp}


\section{Thick brane version of the generalized RS2}
\label{Thick brane version of RS2 generalization}
In this section we will extend the solution found in the previous section for a singular
D3-brane to a thick (smooth) brane scenario in which the thick brane
is dynamically generated by a scalar
field. The action for a 5D scalar field minimally coupled to the Einstein-Hilbert gravity is
\begin{equation}
S=\int dx^5 \sqrt{-g}\left\{2M_\ast^{3}R-\frac{1}{2}g^{MN}\nabla_{M}\phi\nabla_{N}\phi-V(\phi)\right\},
\label{action}
\end{equation}
and we assume the 5D metric to be of the form,
\begin{equation}
ds^2=e^{2A(y)}\eta_{\mu\nu}dx^\mu dx^\nu+dy^2.
\label{G_MN}
\end{equation}
The Einstein equations and the equation of motion for $\phi$, resulting from the action
\eqref{action} are
\begin{eqnarray}
R_{MN}-\frac{1}{2}g_{MN}R&=&\frac{1}{4M_\ast^{3}}T_{MN},\label{eineq1}\\
\nabla^{2}\phi-\frac{dV}{d\phi}&=&0, \label{eineq2}
\end{eqnarray}
where $\nabla^2$ is 5D covariant d'Alembertion operator while the energy-momentum tensor $T_{MN}$
for the scalar field $\phi(y)$ is,
\begin{equation}
T_{MN}=\nabla_{M}\phi\nabla_{N}\phi-g_{MN}\left(\frac{1}{2}(\nabla\phi)^{2}+V(\phi)\right).\label{emt}
\end{equation}
From the Einstein equations \eqref{eineq1} and \eqref{eineq2}, one gets the following equations of motion
for the metric \eqref{G_MN},
\begin{align}
24M_\ast^{3}(A^{\prime})^{2}&=\frac{1}{2}(\phi^{\prime})^{2}-V(\phi),\label{eom01}\\
12M_\ast^{3}A^{\prime\prime}+24M_\ast^{3}(A^{\prime})^{2}&=-\frac{1}{2}(\phi^{\prime})^{2}-V(\phi),\label{eom02}\\
\phi^{\prime\prime}+4A^{\prime}\phi^{\prime}&-\frac{dV}{d\phi}=0. \label{eom03}
\end{align}
We assume that the scalar potential $V(\phi)$ could be expressed
in terms of a superpotential \cite{DeWolfe:1999cp,Kehagias:2000au,Ahmed:2012nh} $W(\phi)$ as follows ,
\begin{equation}
V(\phi)=\frac{1}{2}\left( \frac{\partial W(\phi)}{\partial \phi}\right)^{2}-\frac{1}{6M_\ast^{3}}W(\phi)^{2},
\label{potential}
\end{equation}
where $W(\phi)$ satisfies the following relations,
\begin{equation}
 \phi^{\prime}=\frac{\partial W(\phi)}{\partial \phi} \hspace{1cm}\text{and}\hspace{1cm} A^{\prime}=-\frac{1}{12M_\ast^{3}} W(\phi).
 \label{super_potential}
\end{equation}
Although the use of this method is motivated by supergravity, no supersymmetry is involved in
our set-up. This method is elegant and
very efficient, in particular it reduces the system of second order differential equations
\eqref{eom01}-\eqref{eom03} to first order ordinary differential equations.

We are interested in the case for which the scalar field $\phi(y)$ is given by a kink-like
profile~\footnote{The scalar field $\phi(y)$ profile could be different from the standard kink.
However, as it will be explained in the next section, the profile should be monotonic (invertible)
and $\phi^{\p2}(y)$ should be integrable.}, i.e.,
\begin{equation}
 \phi(y)= \frac{\kappa}{\sqrt{\beta}}\tanh(\beta y),
\label{scalar}
\end{equation}
where $\beta$ is the thickness regulator and $\kappa$ parameterizes tension of the brane in the so called \emph{brane limit}: $\beta\to \infty$.
The energy-density ($T_{00}$) implied by $\phi(y)$ is localized near $y=0$ with the corresponding width controlled by $\beta$.
We will find solutions which mimic a positive-tension brane along with two different cosmological constants on either side of the brane.
If the scalar field $\phi(y)$ is known then the superpotential $W(\phi)$ can be obtained from Eq.~\eqref{super_potential} as,
\begin{align}
\phi^{\prime}(y)&=\frac{\partial W(\phi)}{\partial\phi}=\frac{\partial W(\phi(y))}{\partial y}\frac{\partial y}{\partial\phi(y)}=
\frac{W^{\prime}(y)}{\phi^\prime(y)},\label{super_potential_1}\\
W(y)&=\int_{y_0}^y (\phi^{\prime}(y))^{2}dy+W_{0},
\label{super_potential_2}
\end{align}
where $W_0$ is a constant of integration. It is important to note that in deriving the above relation it is assumed that
$\phi(y)$ must be an invertible
function of $y$, such that $W(\phi)$ can be represented as $W(y)$. Now with the scalar field \eqref{scalar} the superpotential $W(\phi)$ could be explicitly obtained as a function of $y$:
\begin{align}
W(y)=&\kappa^2\left\{\tanh(\beta (y))-\frac{1}{3}\tanh^{3}(\beta (y))\right\}+W_{0}.
\label{super_potential_brane_1}
\end{align}
The integration constant $W_0$ can be fixed by initial conditions imposed upon $A^\p(y)$, e.g. such that
$A'(y_{max})=0)$ for a given $y_{max}$. The non-zero value of $W_0$ turns out to be
essential to reproduce, in the brane limit, the generalized RS2 model presented in the previous section,
whereas for $W_0=0$ the solution for $A(y)$ is symmetric under $y\leftrightarrow -y$ and it corresponds to the standard RS2 in the brane limit \cite{DeWolfe:1999cp,Kehagias:2000au}. It is instructive to write down explicitly
the brane-limit results for the thick brane scenario  in order to determine necessary relations
that must be satisfied to reproduce the RS2 relations (\ref{App_RS}) in the brane limit.
As we will show below there is a direct relation between $W_0\neq0$ and the fact that $\Lambda_+\neq\Lambda_-$.

Let us consider only the scalar field part of the action:
\begin{align}
S_{\phi}&=\int dx^5 \sqrt{-g}\left\{-\frac{1}{2}g^{MN}\nabla_{M}\phi\nabla_{N}\phi-V(\phi)\right\}\notag \\
&=\int dx^5 \sqrt{-g}\left\{-\left(\frac{\partial W(\phi)}{\partial\phi}\right)^{2}+\frac{1}{6M_\ast^{3}}W^{2}(\phi) \right\} \notag \\
&=\int dx^5 \sqrt{-g}\left\{\frac{-\beta\kappa^2}{\cosh^4(\beta ( y))}
+\frac{1}{6M_\ast^{3}}\left[\kappa^2\left(\tanh(\beta ( y))-\frac{1}{3}\tanh^{3}(\beta ( y))\right) +W_{0} \right]^{2}\right\}. \label{action_phi}
\end{align}
In the brane limit, i.e., $\beta\to\infty$ we have,
\[
\lim_{\beta\to\infty}\left\{\frac{\beta}{\cosh^4(\beta ( y))}\right\}=\frac{4}{3}\delta(y),
\]
such that the scalar action \eqref{action_phi} can be written as,
\begin{align}
S_{\phi}
&=\int dx^5 \sqrt{-g}\left\{-\frac{4}{3}\kappa^2\delta(y)-\Lambda_+\Theta(y) - \Lambda_-\Theta(-y) \right\}.  \label{action_phi_bl}
\end{align}
Where $\Lambda_\pm$ are cosmological constants in the bulk for $y\gtrless 0$:
\begin{align}
\Lambda_\pm&=\lim_{\beta\to\infty}\bigg[-\frac{1}{6M_\ast^{3}} \left\{\pm\kappa^2\left(\tanh(\beta ( y))-\frac{1}{3}\tanh^{3}(\beta ( y))\right) +W_{0} \right\}^{2}\bigg],  \notag\\
&=-\frac{1}{6M_\ast^{3}} \left(\pm\frac{2}{3}\kappa^2 +W_{0}\right)^{2}
=-\frac{1}{6 M_\ast^{3}} \left(\frac{\lambda}{2} \pm W_{0}\right)^{2} \lsp y\gtrless 0,
\label{Lambda_pm}
\end{align}
and $\lambda\equiv\frac{4}{3}\kappa^2$ corresponds to the brane tension. Hereafter, we will consider the case
$-\Lambda_+>-\Lambda_-$, that implies $W_0>0$. It is also important to note that Eq.~\eqref{Lambda_pm}
implies that the bulk cosmological constants $\Lambda_\pm$ are negative on either side leading to
anti-de Sitter vacua or in the case with $ W_{0}=\lambda/2$ corresponding to a Minkowski geometry
in that region of space. Equation \eqref{Lambda_pm} implies that in order to reproduce the generalized RS2
scenario defined by a given $M_\ast$, $\lambda$ and $\Lambda_\pm$, the following constraints on the
parameters ($\kappa$, $W_0$) of the thick brane model must hold:
\bea
\kappa^2 &=& \frac34 \lambda \label{kappa}\\
W_0 &=& \sqrt{\frac{3}{2}M_\ast^{3}}\left(\sqrt{-\Lambda_+}-\sqrt{-\Lambda_-}\right)
\label{w0}.
\eea
For consistency of the above choice for $W_0$, the following inequality must hold:
\beq
0 < W_0 < \frac{\lambda}{2}.
\label{wlim1}
\eeq
Therefore, only scenarios with limited splitting between cosmological constants
could be realized:
\beq
\sqrt{6M_\ast^{3}}\left(\sqrt{-\Lambda_+}-\sqrt{-\Lambda_-}\right)<\lambda.
\label{split}
\eeq
Then, for $W_0$ within the limit \eqref{wlim1},  Eq.~\eqref{Lambda_pm} implies that
\beq
\lambda =  \sqrt{6M_\ast^{3}}\left(\sqrt{-\Lambda_+}+\sqrt{-\Lambda_-}\right),
\label{lambda}
\eeq
which is identical as the generalized RS2 relation \eqref{App_RS}.
Note that for the $Z_2$ symmetric case (the standard RS2 model) for which
$\Lambda_+=\Lambda_-=\Lambda_B$, we recover the RS2 relation between the brane
tension and bulk cosmological constant $\lambda=\sqrt{-24M_\ast^{3}\Lambda_B}$
\cite{Randall:1999ee,Randall:1999vf} and $W_0=0$.

It is straightforward to calculate the warp function $A(y)$ by integrating the
second equation in Eq.~\eqref{super_potential} w.r.t. $y$.
The result reads,
\begin{align}
A(y)=&-\frac{\kappa^2}{72M_\ast^{3}\beta}\left(\tanh^{2}(\beta y)+\ln\cosh^{4}(\beta y)\right)-\frac{W_0}{12M_\ast^{3}}y.
\label{warp-factor}
\end{align}
The integration constant above was fixed by the condition $A(0)=0$.
As we have shown in \eqref{w0} $W_0$ is fixed uniquely to a non-zero value, then as a consequence,
in the smooth case the warp function $A(y)$ will not have maxima on the brane location, i.e., $y=0$
but it will be shifted to a position $y_{max}$,
for instance for $M_\ast=1$, $\kappa=1$ and $W_0=0.5M_\ast^3$,
\beq
y_{max} \sim -\frac{0.6}{\beta}.   \label{y_beta}
\eeq
It is worth noticing that even though $A'(0)\neq 0$, nevertheless the maxima of $A(y)$
approaches the brane location, i.e., $y_{max}\to 0$ as $\beta\to \infty$, which is manifested from
the above equation.

%

\begin{figure}[!hbt]\centering
\begin{tabular}{cc}
\includegraphics[scale=0.65]{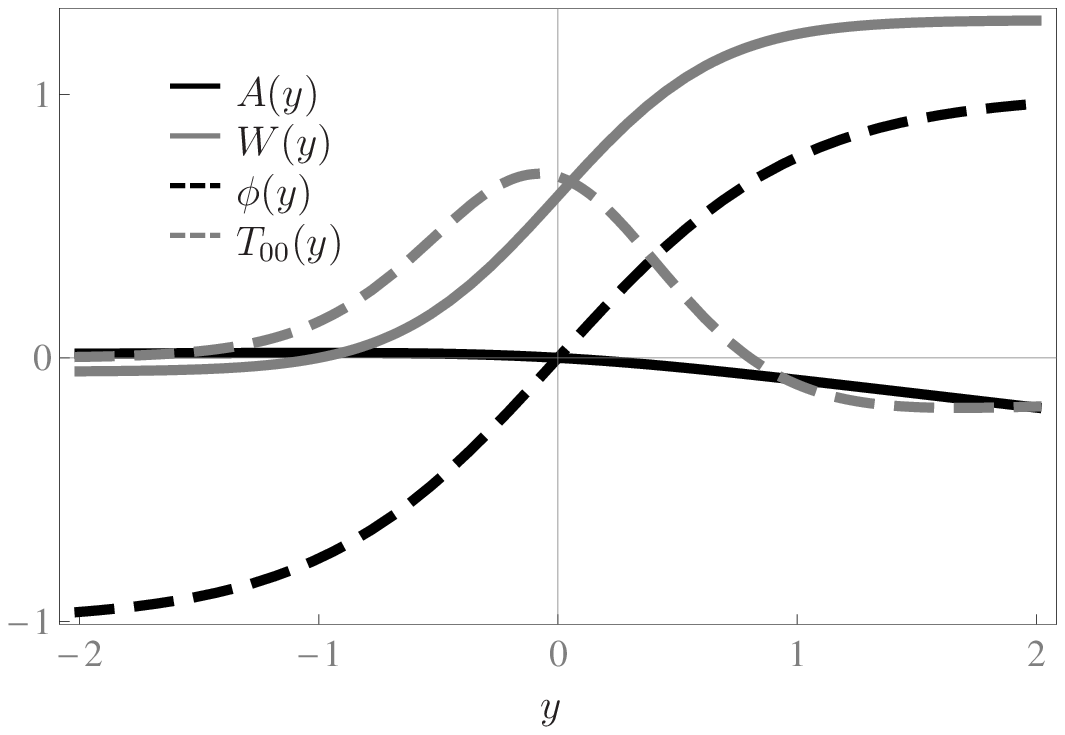} & \includegraphics[scale=0.65]{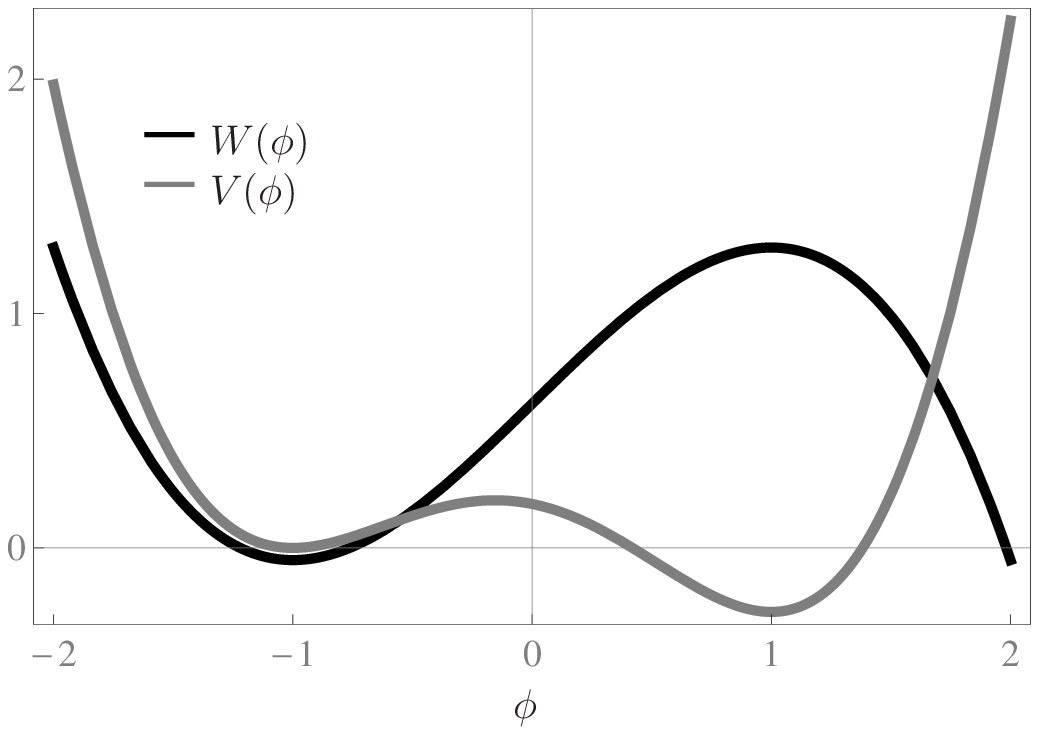}
\end{tabular}
\caption{This left graph shows the behavior of $A(y)$, $W(y)$, $\phi(y)$ and $T_{00}(y)$ as a
function of $y$, whereas, the right graph presents the superpotential $W(\phi)$ and the potential
$V(\phi)$ as a function of the scalar field $\phi$ for $W_0=0.5 M_\ast^4$ and $M_\ast=\beta=\kappa=1$.}
\label{fig_WV}
\end{figure}

Note that far away from the thick brane the warp function approaches the generalized
RS2 form as presented in Sec.~\ref{RS2 generalization},
\begin{align}
A(y)&\approx -k_\pm\vert y\vert, \hspace{2cm} \vert y\vert\to \infty,
\label{A_RS}
\end{align}
where
\[
k_\pm=\frac{1}{24M_\ast^{3}}\lambda\pm\frac{W_0}{12M_\ast^{3}},
\]
It is also important to note that one obtains the same behavior of $A(y)$ \eqref{A_RS},
for all values of $y$ in the brane limit when $\beta\to\infty$, i.e.,
\[
A(y)\approx -k_\pm\vert y\vert, \hspace{2cm} \beta\to \infty \hsp{\rm for}\hsp y\gtrless0.
\]
Since $\phi(y)$ is invertible therefore we can write the superpotential $W(\phi)$ and
the scalar potential $V(\phi)$ as follows:
\bea
W(\phi)&=&\kappa\sqrt{\beta}\phi\left(1-\frac{\beta}{3\kappa^2}\phi^2\right)+W_0,  \label{superpotential}\\
V(\phi)&=&\frac{\beta^3}{2\kappa^2}\left(\phi^2-\frac{\kappa^2}{\beta}\right)^2 -\frac{1}{54M_\ast^{3}}\frac{\beta^3}{\kappa^2}\phi^2\left(\phi^2-3\frac{\kappa^2}{\beta}\right)^2 \non\\
&&+\frac{1}{9M_\ast^{3}}\frac{\beta^{3/2}}{\kappa}\phi\left(\phi^2-3\frac{\kappa^2}{\beta}\right)W_0-\frac{1}{6M_\ast^{3}}W_0^2.
\label{potential_phi}
\eea
Note that the constant term of superpotential $W_0$, in Eq. \eqref{superpotential}, plays the most crucial role in producing the asymmetry in the bulk cosmological constants and then in the warp function $A(y)$ on the left and the right of (thick) brane.
In the left panel of Fig.~\ref{fig_WV} we have shown $y$-dependent shapes of $A(y)$, $W(y)$,
$\phi(y)$ and $T_{00}(y)$,
while in the right one $W(\phi)$ and $V(\phi)$ are plotted as a function of $\phi$.

For the thick brane scenario one can show (following e.g. \cite{Ahmed:2012nh}) that the 4D effective gravity
on the thick brane could be recovered and the background solutions found above are stable.
Here we will only discuss the behavior of the zero mode of tensor perturbations which corresponds
to the 4D graviton and the Schr\"odinger-like potential in the generalized RS2 case with thick brane.

In order to illustrate stability of our solutions for the Einstein equations
let us perturb the metric \eqref{G_MN} such that,
\begin{align}
ds^{2}&=e^{2A(y)}(\eta_{\mu\nu}+H_{\mu\nu})dx^{\mu}dx^{\nu}+dy^{2},
\label{metricT}
\end{align}
where, $H_{\mu\nu}=H_{\mu\nu}(x,y)$ is the transverse and traceless tensor fluctuation, i.e.,
\beq
\partial^\mu H_{\mu\nu}=H^{\mu}_{\mu}=0.
\eeq
One can find the following form of the linearized field equation for the tensor mode,
\begin{align}
\left(\partial_{5}^{2}+4A^{\prime}\partial_{5}+e^{-2A}\Box\right)H_{\mu\nu} &=0,
\label{ten_pert_0}
\end{align}
where $\partial_5\equiv\partial/\partial y$ and $\Box$ is the 4D d'Alembertian operator.
The zero-mode solution (corresponding to $\Box H_{\mu\nu}=0$) of the above equation represents
the 4D graviton while the non-zero modes (corresponding to
$\Box H_{\mu\nu} = m^2 H_{\mu\nu} \neq 0$) are the Kaluza-Klein (KK) graviton excitations.

In order to gain more intuition and understanding of the tensor mode equation of motion \eqref{ten_pert_0},
it is convenient to change
the variables such that we can get rid of the exponential factor in front of the d'Alembertian
and the single derivative term with
$A^{\prime}$, so that we convert the above equation into the standard Schr\"odinger like form.
We can achieve this in two
steps; first by changing coordinates such that the metric becomes conformally flat:
\begin{align}
ds^{2}&=e^{2A(z)}\left(\eta_{\mu\nu}dx^{\mu}dx^{\nu}+dz^{2}\right),
\label{metric_1}
\end{align}
with $z$ defined through the differential equation: $dz=e^{-A(y)}dy$. In the new coordinates
the Eq.~\eqref{ten_pert_0} takes the form
\begin{align}
\left(\partial_{z}^{2}+3\dot{A}(z)\partial_{z}+\Box\right)H_{\mu\nu} &=0, \label{ten_pert_1}
\end{align}
where $dot$ over $A$ represents a derivative with respect to $z$ coordinate. Now we can
perform the second step
removing the single derivative term in \eqref{ten_pert_1} by the following
redefinition of the tensor fluctuation
\begin{align}
H_{\mu\nu}(x,z) &=e^{-3A(z)/2}\tilde{H}_{\mu\nu}(x,z).
\label{tilde_H}
\end{align}
Hence the Eq.~\eqref{ten_pert_1} will take the form of the Schr\"odinger equation,
\begin{align}
\left(\partial_{z}^{2}-\frac{9}{4}\dot{A}^{2}(z)-\frac{3}{2}\ddot{A}(z)+\Box\right)\tilde{H}_{\mu\nu}(x,z) &=0.
\label{schrodinger_eq_0}
\end{align}
We can decompose the $\tilde H_{\mu\nu}(x,z)$ into the $x$ and $z$ dependent parts as
$\tilde H_{\mu\nu}(x,z)=\hat H_{\mu\nu}(x)\psi(z)$. Where $\hat H_{\mu\nu}(x)\propto e^{ipx}$
is a $z$-independent plane wave solution such that $\Box \hat H_{\mu\nu}(x)=m^2\hat H_{\mu\nu}(x)$,
with $-p^2=m^2$ being the 4D KK mass of the tensor mode. Then the above equation takes the form,
\beq
\bigg(-\partial_{z}^{2}+{\cal V}(z)\bigg)\psi(z) = m^{2}\psi(z), \label{schrodinger_eq}
\eeq
where ${\cal V}(z)$ is the Schr\"odinger-like potential,
\beq
{\cal V}(z) =\frac{9}{4}\dot{A}^{2}(z)+\frac{3}{2}\ddot{A}(z).
\label{V_potential}
\eeq
Note that we can rewrite the Schr\"odinger-like equation \eqref{schrodinger_eq} in supersymmetric
quantum mechanics form as,
\beq
{\cal Q}^{\dagger}{\cal Q}\psi  =\left(-\partial_{z}-\frac{3}{2}\dot{A}\right)\left(\partial_{z}-\frac{3}{2}\dot{A}\right)\psi
=m^{2}\psi.
\label{susy_qm_0}
\eeq
The zero mode ($m^2=0$) profile, $\psi_0(z)$, corresponds to the graviton in the 4D effective theory.
The stability
with respect to the tensor fluctuations of the background solution is guaranteed by the positivity
of the operator
${\cal Q}^{\dagger}{\cal Q}$ in the supersymmetric quantum mechanics version of the equation of
motion \eqref{susy_qm_0} as it forbids the existence of any tachyonic mode with negative mass square,
$m^{2}<0$~\footnote{Since $\int dz ({\cal Q} \psi)^2+\psi {\cal Q} \psi\left|_{-\infty}^{+\infty}\right. = m^2 \int dz \psi^2$
and the first term $\int dz ({\cal Q} \psi)^2$ is definite non-negative, therefore in order to
guarantee $m^2\geq0$ the
boundary term (second term) must vanish or be positive.}.
So, in that case, the perturbation is not growing in time, hence the background solution is stable.

\begin{figure}[!hbt]\centering
\includegraphics[scale=0.7]{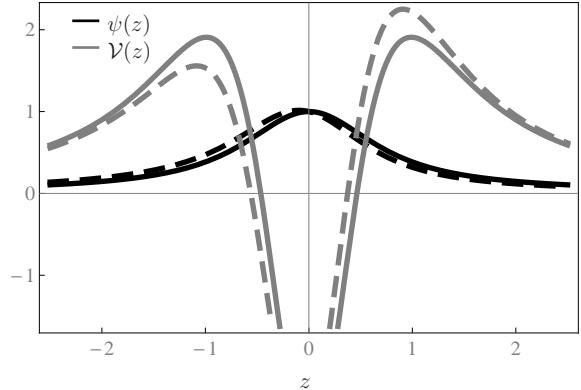}
\caption{This graph shows the behavior of the zero-mode for tensor perturbations $\psi(z)$ and the Schr\"odinger-like
potential ${\cal V}(z)$ as a
function of $z$. The solid lines correspond to the symmetric case $W_0=0$, whereas, the dashed lines refer to the
asymmetric case with $W_0=0.5 M_\ast^4$ for $M_\ast=\beta=1$ and $\kappa=5$.}
\label{fig_psi_V}
\end{figure}

The zero-mode wave function $\psi_0(z)$ can be obtained by noticing that
\begin{align}
{\cal Q}\psi_0 &=\left(\partial_{z}-\frac{3}{2}\dot{A}\right)\psi_0=0,
\label{susy_qm_1}
\end{align}
which implies that,
\begin{align}
\psi_0(z) &=e^{\frac{3}{2}A(z)} .
\label{bar_H_0}
\end{align}
In Fig. \ref{fig_psi_V} we have plotted the zero mode of tensor perturbations $\psi_0(z)=e^{\frac{3}{2}A(z)}$ given by Eq. \eqref{bar_H_0} and ${\cal V}(z)$ for the warp-function $A[y(z)]$ Eq. \eqref{warp-factor}.
One can make the following comments resulting from the profile of the zero-mode for tensor perturbations $\psi_0(z)$ and the
Schr\"odinger-like potential ${\cal V}(z)$ shown in Fig.~\ref{fig_psi_V}:
\bit
\item The zero-mode $\psi_0(z)$ implies that
\begin{align}
\int dz \psi_0^{2}(z) = \int dz e^{3A(z)} = \int dy e^{2A(y)} < \infty,
\label{normalizablity_y}
\end{align}
therefore $\psi_0(z)$ is normalizable and it turns out that the effective 4D Planck mass $M^{2}_{Pl}$ is finite, hence the effective 4D gravity can be reproduced for the thick brane case.
\item As ${\cal V}(z)\to 0$ as $|z|\to\infty$, therefore the KK-mass spectrum is continuous
without a gap and it starts from $m=0$.
\item The (asymmetric) volcano-like shape of ${\cal V}(z)$ in Fig.~\ref{fig_psi_V}
suggests that at large $z$ the wave function massive KK modes should have a plane wave behaviour.
\item The presence of the large barriers near the thick brane (z=0) implies that corrections to the
Newton's law due to continuum spectrum of the KK modes will not be large \cite{Csaki:2000fc,Csaki:2000pp}.
\eit


\section{Generalized thick branes}
\label{Generalized thick branes}
In this section we will consider a general case for the background scalar field.
We are going to show that even without a priori defined shape of the scalar field profile,
the thin brane generalized RS2 relation (\ref{App_RS}) between the brane tension $\lambda$ and the bulk
cosmological constants $\Lambda_\pm$ is reproduced in the brane limit under certain mild assumptions.
In other words the relation is independent of the function adopted to regularize (smooth)
a thin brane. For this purpose we consider the following general form of the scalar background field,
\beq
\phi (y) = \frac{\phi_0 (\beta y)}{\sqrt{\beta}} , \label{phi_general}
\eeq
where $\beta$ will turn out to be the thickness controlling parameter.
{\it We assume that $\phi_0 (\beta y)$ is monotonic, and $(\sqrt{\beta}\phi_0' (\beta y))^2$
is an integrable function of $y$}~\footnote{It is interesting to notice that this condition is
equivalent to the normalizability of one of the two scalar zero modes (spin zero fluctuations
around the background solution \eqref{scalar} and \eqref{warp-factor}) related
to the shift along the extra dimension $y \to y + {\rm const.}$, for more details see \cite{Ahmed:2012nh}.}.
We use the superpotential method described in the previous section. It is worth to note here that the
method is equivalent to the standard approach (i.e. solving the Einstein equations) as long as the solutions for scalar field have monotonic profile. Let us consider the scalar field action
\begin{align}
S_\phi &=\int dx^5 \sqrt{-g}\left\{-\frac{1}{2}g^{MN}\nabla_{M}\phi\nabla_{N}\phi-V(\phi)\right\}\notag \\
&=\int dx^5 \sqrt{-g}\left\{-\left(\frac{\partial W(\phi)}{\partial\phi}\right)^{2}+\frac{1}{6M_\ast^{3}}W^{2}(\phi) \right\} \notag \\
&=\int d^5 x  \sqrt{- g} \left\{ - (\phi')^2 + \frac{1}{6 M_\ast^{3}} \left( \int_0^y (\phi'( \bar y))^2 d\bar y + W_0 \right)^2 \right\},
\end{align}
where $V(\phi)$ and $W(\phi)$ are obtained from Eqs. \eqref{potential} and \eqref{super_potential}, respectively.
Since $W_0$ is an arbitrary integration constant the lower integration limit could be chosen at $\bar y=0$
without compromising generality. After using equation \eqref{phi_general} and changing variables
from $\tilde y\to\beta \bar y$ one gets
\beq
S_\phi = \int d^5x \sqrt{- g} \left\{ - \beta (\phi_0' (\beta y))^2 + \frac{1}{6 M_\ast^{3}} \left( \int_0^{\beta y} (\phi_0' (\tilde y))^2 d\tilde y + W_0 \right)^2 \right\} . \label{Sphi_general}
\eeq
From the above scalar field action, one finds that in the brane limit,
i.e., $\beta \rightarrow \infty$:
\bit
\item
The integrand $\beta (\phi_0' (\beta y))^2$ converges to zero everywhere
except $y=0$ (as the function is integrable) therefore the first term above approaches
$- \lambda \delta(y)$, with
\[
\lambda = \int_{- \infty}^{+ \infty} (\phi_0' (\tilde y))^2 d\tilde y,
\]
where $\delta(y)$ is the Dirac delta function.
\item
The second term converges to a sum of contributions to bulk cosmological constants $- \Lambda_+\Theta(y) - \Lambda_-\Theta(-y)$, where
\bea
\Lambda_+ &=& -\frac{1}{6M_\ast^{3}}\left( \int_0^{+\infty} (\phi_0'(\tilde y))^2 d\tilde y + W_0 \right)^2 \label{Lam_+}\\
\Lambda_- &=& -\frac{1}{6M_\ast^{3}}\left( -\int_{-\infty}^0 (\phi_0'(\tilde y))^2 d\tilde y + W_0 \right)^2. \label{Lam_-}
\eea
\eit
Equations~\eqref{Lam_+}-\eqref{Lam_-} imply that in order to reproduce the generalized RS2 relation
(\ref{App_RS}) the following inequality must hold
\beq
\sqrt{6M_\ast^{3}}\left(\sqrt{-\Lambda_+}-\sqrt{-\Lambda_-} \right) < \lambda.
\label{split2}
\eeq
Note that this is the same condition that was limiting the splitting between
the cosmological constants which was obtained in Sec.~\ref{Thick brane version of RS2 generalization}.
Therefore we conclude that regardless what is the choice of the scalars profile,
only those thin brane models could be obtained in the brane limit for which
\eqref{split2} is satisfied.

It is easy to see that if $W_0$ is chosen as
\beq
W_0=\sqrt{\frac32 M_\ast^{3}}\left(\sqrt{-\Lambda_+}-\sqrt{-\Lambda_-} \right)
+ \frac12 \left( \int_{-\infty}^0 (\phi_0' (y))^2 dy - \int_0^{+\infty} (\phi_0' (y))^2 dy \right),
\label{w02}
\eeq
then indeed
\beq
\lambda = \sqrt{6M_\ast^{3}}\left(\sqrt{-\Lambda_+}+\sqrt{-\Lambda_-} \right).
\label{genrel}
\eeq
Thus we recover the result \eqref{App_RS} for our generalized RS2 model.
It is worth to rephrase the above result as follows. For any
given thin brane model to be reproduced in the brane limit and any profile
of the scalar field $\phi_0(y)$ (monotonic with $(\phi_0'(y))^2$ integrable),
the Eq.~\eqref{w02} provides the choice of the integration constant $W_0$
which guaranties that the condition \eqref{App_RS} holds.

In the case of the kink-like profile considered in Sec.~\ref{Thick brane version of RS2 generalization},
$(\phi_0' (y))^2$ was an even function of $y$ therefore $W_0$ reduces to the value
adopted in \eqref{w0}. Of course, if we limit ourself to the $Z_2$-symmetric case,
$W_0$ must vanish as in \cite{DeWolfe:1999cp}.

\section{Summary}
\label{sum}
We have discussed a thick-brane version of the Randall-Sundrum model 2 in which we allow
for different cosmological constants on two sides of the brane. Einstein equations
have been solved and stability of the solution has been illustrated.
The thin brane limit of the model have been discussed.
Properties of the thick brane solution have been considered in details. It has been shown that,
under mild assumptions, the relation between cosmological constants and the brane tension
of the Randall-Sundrum model 2 could be obtained in the brane limit of our model by an appropriate
choice of an integrating constant (that defines the scalar potential) independently
of particular profile of the scalar field.\\


\noindent {\bf Note added:} After this paper has appeared, another interesting study on the same subject has been publicized in Ref. \cite{Bazeia:2013usa}.

\section*{Acknowledgments}
We are grateful to the NORDITA Program ``Beyond the LHC'' for hospitality
during the early stage of this work.
This work has been supported in part by the National Science Centre (Poland)
as a research project, decision no DEC-2011/01/B/ST2/00438.
AA acknowledges financial support from the Foundation for Polish Science
International PhD Projects Programme co-financed by the EU European
Regional Development Fund.

\providecommand{\href}[2]{#2}\begingroup\raggedright
\endgroup


\begin{thebibliography}{10}

\bibitem{Randall:1999ee}
L.~Randall and R.~Sundrum, {\it {A Large mass hierarchy from a small extra
  dimension}},  {\em Phys.Rev.Lett.} {\bf 83} (1999) 3370--3373,
  [\href{http://xxx.lanl.gov/abs/hep-ph/9905221}{{\tt hep-ph/9905221}}].

\bibitem{Randall:1999vf}
L.~Randall and R.~Sundrum, {\it {An Alternative to compactification}},  {\em
  Phys.Rev.Lett.} {\bf 83} (1999) 4690--4693,
  [\href{http://xxx.lanl.gov/abs/hep-th/9906064}{{\tt hep-th/9906064}}].

\bibitem{Gibbons:2000tf}
G.~W. Gibbons, R.~Kallosh, and A.~D. Linde, {\it {Brane world sum rules}},
  {\em JHEP} {\bf 0101} (2001) 022,
  [\href{http://xxx.lanl.gov/abs/hep-th/0011225}{{\tt hep-th/0011225}}].

\bibitem{Ahmed:2012nh}
A.~Ahmed and B.~Grzadkowski, {\it {Brane modeling in warped extra-dimension}},
  {\em JHEP} {\bf 1301} (2013) 177,
  [\href{http://xxx.lanl.gov/abs/1210.6708}{{\tt arXiv:1210.6708}}].

\bibitem{DeWolfe:1999cp}
O.~DeWolfe, D.~Freedman, S.~Gubser, and A.~Karch, {\it {Modeling the
  fifth-dimension with scalars and gravity}},  {\em Phys.Rev.} {\bf D62} (2000)
  046008, [\href{http://xxx.lanl.gov/abs/hep-th/9909134}{{\tt
  hep-th/9909134}}].

\bibitem{Kehagias:2000au}
A.~Kehagias and K.~Tamvakis, {\it {Localized gravitons, gauge bosons and chiral
  fermions in smooth spaces generated by a bounce}},  {\em Phys.Lett.} {\bf
  B504} (2001) 38--46, [\href{http://xxx.lanl.gov/abs/hep-th/0010112}{{\tt
  hep-th/0010112}}].

\bibitem{Rubakov:1983bb}
V.~Rubakov and M.~Shaposhnikov, {\it {Do We Live Inside a Domain Wall?}},  {\em
  Phys.Lett.} {\bf B125} (1983) 136--138.

\bibitem{Gremm:1999pj}
M.~Gremm, {\it {Four-dimensional gravity on a thick domain wall}},  {\em
  Phys.Lett.} {\bf B478} (2000) 434--438,
  [\href{http://xxx.lanl.gov/abs/hep-th/9912060}{{\tt hep-th/9912060}}].

\bibitem{Csaki:2000fc}
C.~Csaki, J.~Erlich, T.~J. Hollowood, and Y.~Shirman, {\it {Universal aspects
  of gravity localized on thick branes}},  {\em Nucl.Phys.} {\bf B581} (2000)
  309--338, [\href{http://xxx.lanl.gov/abs/hep-th/0001033}{{\tt
  hep-th/0001033}}].

\bibitem{Kobayashi:2001jd}
S.~Kobayashi, K.~Koyama, and J.~Soda, {\it {Thick brane worlds and their
  stability}},  {\em Phys.Rev.} {\bf D65} (2002) 064014,
  [\href{http://xxx.lanl.gov/abs/hep-th/0107025}{{\tt hep-th/0107025}}].

\bibitem{Melfo:2002wd}
A.~Melfo, N.~Pantoja, and A.~Skirzewski, {\it {Thick domain wall space-times
  with and without reflection symmetry}},  {\em Phys.Rev.} {\bf D67} (2003)
  105003, [\href{http://xxx.lanl.gov/abs/gr-qc/0211081}{{\tt gr-qc/0211081}}].

\bibitem{Bronnikov:2003gg}
K.~A. Bronnikov and B.~E. Meierovich, {\it {A General thick brane supported by
  a scalar field}},  {\em Grav.Cosmol.} {\bf 9} (2003) 313--318,
  [\href{http://xxx.lanl.gov/abs/gr-qc/0402030}{{\tt gr-qc/0402030}}].

\bibitem{Bazeia:2003aw}
D.~Bazeia, C.~Furtado, and A.~Gomes, {\it {Brane structure from scalar field in
  warped space-time}},  {\em JCAP} {\bf 0402} (2004) 002,
  [\href{http://xxx.lanl.gov/abs/hep-th/0308034}{{\tt hep-th/0308034}}].

\bibitem{CastilloFelisola:2004eg}
O.~Castillo-Felisola, A.~Melfo, N.~Pantoja, and A.~Ramirez, {\it {Localizing
  gravity on exotic thick three-branes}},  {\em Phys.Rev.} {\bf D70} (2004)
  104029, [\href{http://xxx.lanl.gov/abs/hep-th/0404083}{{\tt
  hep-th/0404083}}].

\bibitem{Guerrero:2005aw}
R.~Guerrero, R.~O. Rodriguez, and R.~S. Torrealba, {\it {De-Sitter and double
  asymmetric brane worlds}},  {\em Phys.Rev.} {\bf D72} (2005) 124012,
  [\href{http://xxx.lanl.gov/abs/hep-th/0510023}{{\tt hep-th/0510023}}].

\bibitem{Farakos:2007ua}
K.~Farakos, G.~Koutsoumbas, and P.~Pasipoularides, {\it {Graviton localization
  and Newton's law for brane models with a non-minimally coupled bulk scalar
  field}},  {\em Phys.Rev.} {\bf D76} (2007) 064025,
  [\href{http://xxx.lanl.gov/abs/0705.2364}{{\tt arXiv:0705.2364}}].

\bibitem{Bazeia:2008zx}
D.~Bazeia, A.~Gomes, L.~Losano, and R.~Menezes, {\it {Braneworld Models of
  Scalar Fields with Generalized Dynamics}},  {\em Phys.Lett.} {\bf B671}
  (2009) 402--410, [\href{http://xxx.lanl.gov/abs/0808.1815}{{\tt
  arXiv:0808.1815}}].

\bibitem{Guo:2010az}
H.~Guo, Y.-X. Liu, S.-W. Wei, and C.-E. Fu, {\it {Gravity Localization and
  Effective Newtonian Potential for Bent Thick Branes}},  {\em Europhys.Lett.}
  {\bf 97} (2012) 60003, [\href{http://xxx.lanl.gov/abs/1008.3686}{{\tt
  arXiv:1008.3686}}].

\bibitem{Dzhunushaliev:2009va}
V.~Dzhunushaliev, V.~Folomeev, and M.~Minamitsuji, {\it {Thick brane
  solutions}},  {\em Rept.Prog.Phys.} {\bf 73} (2010) 066901,
  [\href{http://xxx.lanl.gov/abs/0904.1775}{{\tt arXiv:0904.1775}}].

\bibitem{Gabadadze:2006jm}
G.~Gabadadze, L.~Grisa, and Y.~Shang, {\it {Resonance in asymmetric warped
  geometry}},  {\em JHEP} {\bf 0608} (2006) 033,
  [\href{http://xxx.lanl.gov/abs/hep-th/0604218}{{\tt hep-th/0604218}}].

\bibitem{Gregory:2000jc}
R.~Gregory, V.~Rubakov, and S.~M. Sibiryakov, {\it {Opening up extra dimensions
  at ultra large scales}},  {\em Phys.Rev.Lett.} {\bf 84} (2000) 5928--5931,
  [\href{http://xxx.lanl.gov/abs/hep-th/0002072}{{\tt hep-th/0002072}}].

\bibitem{Dvali:2000hr}
G.~Dvali, G.~Gabadadze, and M.~Porrati, {\it {4-D gravity on a brane in 5-D
  Minkowski space}},  {\em Phys.Lett.} {\bf B485} (2000) 208--214,
  [\href{http://xxx.lanl.gov/abs/hep-th/0005016}{{\tt hep-th/0005016}}].

\bibitem{Dvali:2000rv}
G.~Dvali, G.~Gabadadze, and M.~Porrati, {\it {Metastable gravitons and infinite
  volume extra dimensions}},  {\em Phys.Lett.} {\bf B484} (2000) 112--118,
  [\href{http://xxx.lanl.gov/abs/hep-th/0002190}{{\tt hep-th/0002190}}].

\bibitem{Csaki:2000pp}
C.~Csaki, J.~Erlich, and T.~J. Hollowood, {\it {Quasilocalization of gravity by
  resonant modes}},  {\em Phys.Rev.Lett.} {\bf 84} (2000) 5932--5935,
  [\href{http://xxx.lanl.gov/abs/hep-th/0002161}{{\tt hep-th/0002161}}].

\bibitem{Bazeia:2013usa}
D.~Bazeia, R.~Menezes, and R.~da~Rocha, {\it {A Note on Asymmetric Thick
  Branes}},  \href{http://xxx.lanl.gov/abs/1312.3864}{{\tt arXiv:1312.3864}}.

\end{thebibliography}
\end{document}